\documentstyle[twocolumn,epsf]{jpsj}

\catcode`\@=11
\def\simle{\mathrel{\mathpalette\@versim<}}   % < over \sim
\def\simge{\mathrel{\mathpalette\@versim>}}   % > over \sim
\def\@versim#1#2{\lower2.5pt\vbox{\baselineskip0pt \lineskip-.5pt
   \ialign{$\m@th#1\hfil##\hfil$\crcr#2\crcr\sim\crcr}}}

\catcode`\@=12

\def\runtitle{
}
\def\runauthor{Yukitoshi {\sc Motome} and Nobuo {\sc Furukawa}}

\def\sf{\rm}

\title{
Spectral Function Analysis on Spin Dynamics in\\
Double-Exchange Systems with Randomness
}

\author{
Yukitoshi {\sc Motome} and Nobuo {\sc Furukawa}$^{1}$ 
}
\inst{
Tokura Spin SuperStructure Project (SSS), ERATO,
Japan Science and Technology Corporation (JST),
c/o National Institute of Advanced Industrial Science and Technology (AIST),
Tsukuba, Ibaraki 305-8562, Japan\\
$^{1}$Department of Physics, Aoyama Gakuin University, 
Setagaya, Tokyo 157-8572
}
\recdate{October 22,2002}

\abst{
Spin excitation spectrum
is studied in the double-exchange model with randomness.
Applying the spin wave approximation and the spectral function analysis,
we examine excitation energy and linewidth
using analytical as well as numerical methods.
For small wave number $q \sim 0$,
the excitation energy is cosine-like
and the linewidth shows a $q$-linear behavior.
This indicates that the spin excitation becomes
incoherent or localized near $q = 0$.
Crossover takes place to marginally-coherent regime
where both the excitation energy and the linewidth 
are proportional to $q^2$.
The incoherence is due to local fluctuations
of the kinetic energy of electrons.
Comparison with experimental results
in colossal magnetoresistance manganites suggests that
spatially-correlated or mesoscopic-scale fluctuations
are more important in real compounds
than local or atomic-scale ones.
}

\kword{
colossal magnetoresistance manganites,
double-exchange model, randomness, spin wave excitation,
linewidth
}

\begin{document}
\sloppy
\maketitle

%%%%% Introduction %%%%%
\noindent
{\it Introduction} ---
One of main issues
in colossal magnetoresistance (CMR) manganites AMnO$_3$
is whether the double-exchange (DE) mechanism is enough or not,
and, if not, what is necessary as an additional mechanism.
\cite{Furukawa1999}.
The metallic ferromagnetism is well described
by the DE mechanism qualitatively
\cite{Zener1951},
which indicates that
the DE interaction is obviously an essential element
in these materials.
Recently, however,
many experimental aspects have been indicated 
the necessity of additional elements beyond it.
For instance, insulating behavior above the Curie temperature $T_{\rm C}$
as well as large residual resistivity is observed
in compounds, such as (La,Ca)MnO$_3$,
which show relatively low $T_{\rm C}$
\cite{Schiffer1995}.
The DE model predicts metallic conductivity 
both above and below $T_{\rm C}$
as observed in compounds which have high $T_{\rm C}$,
such as (La,Sr)MnO$_3$
\cite{Urushibara1995}.
Thus, another mechanism might be necessary
to describe the former `low-$T_{\rm c}$' compounds.
Since these deviations from the canonical DE behavior
appear to be systematic for the A-site substitution
\cite{Saitoh1999},
the question is what is the controlling parameter.
The magnitude of CMR effects becomes larger
in lower-$T_{\rm C}$ materials,
therefore the issue attracts much attentions
from the viewpoint of the basic mechanism of CMR.

One of crucial tests for an additional mechanism 
is spin dynamics.
Spin excitation spectrum shows different behavior
between high-$T_{\rm C}$ and low-$T_{\rm C}$ manganites.
In high-$T_{\rm C}$ compounds, 
the spin excitation shows a cosine-like dispersion
\cite{Perring1996},
which is well reproduced by the DE mechanism alone
\cite{Furukawa1996}.
On the contrary, in low-$T_{\rm C}$ compounds,
the spectrum deviates from this form
and exhibits some anomalies 
such as broadening, softening and gap-opening
\cite{Hwang1998,Vasiliu-Doloc1998,Dai2000,Biotteau2001}.
Several mechanisms have been proposed to explain 
these anomalies
\cite{Furukawa1999a,Solovyev1999,Khaliullin2000,Golosov2000,Shannon2002,Motome2002}.
Most of them can reproduce qualitative features of
the anomalous spin excitations to some extent,
hence, 
more quantitative comparison between experimental and theoretical results
is desired to determine an essential mechanism.

An intriguing point is that 
spin excitations in low-$T_{\rm C}$ manganites show large intrinsic
linewidths even at the lowest temperature
\cite{Hwang1998,Vasiliu-Doloc1998,Dai2000,Biotteau2001,Perring2001}.
This indicates that magnons in the ground state
are not the eigenstates of the system.
Theoretical models which can be effectively mapped to
Heisenberg spin systems with uniform exchange couplings
cannot describe this situation.

The authors have claimed that the anomalous spin excitation
is well reproduced by introducing the randomness 
\cite{Motome2002}.
Randomness appears to be promising among many proposed scenario
since it is inherently controlled by the A-site substitution and
gives a comprehensive understanding for 
systematic changes of the spin excitation 
from the high-$T_{\rm C}$ to the low-$T_{\rm C}$ materials,
at least, in a qualitative level.
In the presence of randomness,
magnons are no longer the eigenstates, and
the spin excitation shows 
a deviation from the cosine-like dispersion and a finite linewidth.
It is strongly desired to compare the excitation spectrum
including the linewidth quantitatively
with experimental results.

In this Letter, we study the randomness effect on 
the spin wave spectrum quantitatively,
especially on
the excitation energy and the linewidth,
within the spin wave approximation.
We find an incoherent magnon excitation 
in the vicinity of the zone center and
a crossover to a marginally-coherent regime
where the linewidth is almost linearly scaled
to the excitation energy.

%%%%% Model %%%%%
\noindent
{\it Formulation} ---
We investigate spin dynamics in the DE model with quenched randomness.
The Hamiltonian includes both diagonal and off-diagonal disorder,
which is explicitly given by
\begin{equation}
{\cal H} 
=
-\sum_{ij, \sigma} 
t_{ij}
(c_{i \sigma}^\dagger c_{j \sigma} + {\rm h.c.})
- J_{\rm H} \sum_i \mib{\sigma}_i \cdot \mib{S}_i
+ \sum_{i \sigma} 
\varepsilon_i
c_{i \sigma}^{\dagger} c_{i \sigma},
\label{eq:H}
\end{equation}
where the first term denotes the electron hopping,
the second one is the Hund's-rule coupling 
between itinerant electrons 
and localized spins with a magnitude $S$, and
the last one gives the on-site potential.
The diagonal and off-diagonal disorder are incorporated
in the on-site potential $\varepsilon_i$ and
the transfer integral $t_{ij}$, respectively.
We call the former `the on-site randomness' and
the latter `the bond randomness' hereafter.
In the following, we first discuss these randomness effects
on spin dynamics in a general formulation
which does not depend on either details of the type of randomness
or the dimension of the system.
Later, we compare the analytic results with numerical ones.

By applying the spin wave approximation
in the lowest order of $1/S$ expansion,
the spin wave excitations 
in the limit of $J_{\rm H}/t \rightarrow \infty$
are obtained from
the static part of the magnon self-energy
\cite{Motome2002,Furukawa1996},
\begin{equation}
\Pi_{ij}
=
\frac{1}{2S} \sum_{mn} f_{n+}
 \varphi_{n+}(j) \varphi_{n+}^* (i)
\varphi_{m-}(i) \varphi_{m-}^* (j)
(E_{m} - E_{n}),
\label{eq:Pi_org}
\end{equation}
where 
$f_{n+}$ is the fermi distribution function for up-spin states.
Here $\varphi_{n\sigma}(i)$ is 
the $n$-th orthonormal eigenfunction, 
which satisfies
$
\sum_j {\cal H}_{ij} (\{t_{ij},\varepsilon_{i} \}) 
\varphi_{n\sigma}(j) 
= (E_n - \sigma J_{\rm H}) \varphi_{n\sigma}(i)
$
for Hamiltonian (\ref{eq:H}) with a given configuration of randomness.
The spectral function for the spin wave, $A(\mib{q},\omega)$,
is calculated by averaging the quantity
\begin{equation}
A(\mib{q},\omega) = \frac{1}{N}
\sum_l \Bigl| \sum_{j}
\psi_{l}(j) {\rm e}^{{\rm i}\mib{q}\mib{r}_{j}} \Bigr|^2
\delta(\omega - \omega_{l})
\label{eq:A_qw}
\end{equation}
for random configurations.
Here, $N$ is the system size;
$\omega_{l}$ and $\psi_{l}(j)$ are 
the eigenvalues and eigenfunctions
of $\Pi_{ij}$, respectively, which satisfy
\begin{equation}
\sum_j \Pi_{ij} \psi_{l}(j)
 = \omega_{l} \psi_{l}(i).
\label{eq:eigeneq_Pi}
\end{equation}
Note that
the observable quantities are to be averaged finally
for configurations of randomness.

Before going into the analysis on the spectral function,
we here discuss the magnon self-energy $\Pi_{ij}$ in detail.
Eq.~(\ref{eq:Pi_org}) can be written in the form
\begin{equation}
\Pi_{ij} =
\frac{1}{2S} \big(
{\cal H}_{ij} B_{ji}  - \delta_{ij}
\sum_k {\cal H}_{ik} B_{ki} \big),
\label{eq:Pi}
\end{equation}
when we define
$
B_{ji} = \sum_n f_n \varphi_n (j) \varphi_n^* (i)
$
and use the relations
$
\sum_m E_m \varphi_m (i) \varphi_m^* (j)
=
\sum_{mk} {\cal H}_{ik} \varphi_m (k) \varphi_m^* (j)
= {\cal H}_{ij}
$
and
$
\sum_n f_n E_n \varphi_n (j) \varphi_n^* (i)
=
\sum_k {\cal H}_{jk} B_{ki}
$
with the orthonormal property of $\varphi_n (i)$.
Here we drop the spin indices for simplicity.
Apparently from eq.~(\ref{eq:Pi}),
$\Pi_{ij}$ satisfies the sum rule
\begin{equation}
\sum_j \Pi_{ij} = 0.
\label{eq:sumrule}
\end{equation}
Moreover, the matrix element $\Pi_{ij}$ 
consists of the transfer energy of electrons as
\begin{eqnarray}
\Pi_{i\neq j} &=&
\frac{1}{2S} {\cal H}_{ij} B_{ji}
= -\frac{1}{2S} t_{ij} \langle c_i^\dagger c_j \rangle
\equiv - 2S J_{ij},
\nonumber \\
\Pi_{ii} &=&
- \frac{1}{2S} \sum_{j\neq i} {\cal H}_{ij} B_{ji} =
2S \sum_j J_{ij},
\label{eq:Pi_ij}
\end{eqnarray}
where 
$
J_{ij} = t_{ij} \langle c_i^\dagger c_j \rangle / 4S^2
$
is the exchange coupling of the corresponding Heisenberg model 
within the lowest order of the spin wave expansion.
\cite{FurukawaPreprint}
The bracket denotes the expectation value in the ground state
for a given configuration of randomness.
Hence the following summations equal to
the kinetic energy of electrons as
\begin{equation}
\sum_i \Pi_{ii} 
= - \sum_{i \neq j} \Pi_{ij}
= - \langle T \rangle / 2S,
\label{eq:Pi_ii}
\end{equation}
where $T$ is the first term in the Hamiltonian (\ref{eq:H}).

\begin{figure}
\epsfxsize=8.5cm
\centerline{\epsfbox{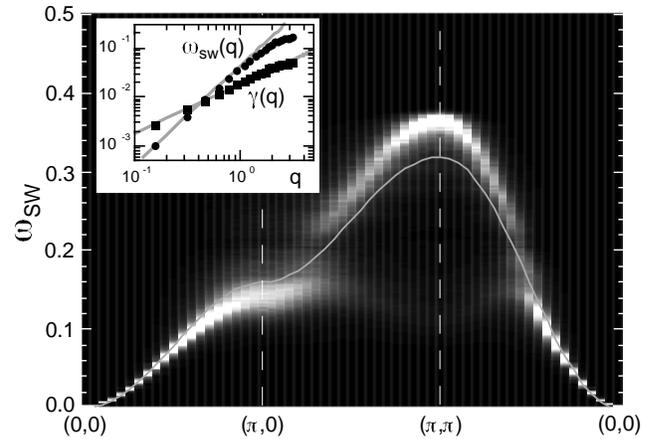}}
\caption{
Spin excitation spectrum in the case of
the on-site randomness
$\delta\varepsilon = 0.25$.
The gray curve is the excitation energy 
calculated by eq.~(\protect{\ref{eq:epsilon_q}}).
Inset: The excitation energy $\omega_{\rm sw}$ and 
the linewidth $\gamma$
as a function of $(q,0)$ for $0 < q < \pi$.
The lines are the fits by $q^2$ and $q$
for $\omega_{\rm sw}$ and $\gamma$, respectively.
}
\label{fig:disp}
\end{figure}

Now we apply the spectral function analysis 
on the spin excitation spectrum.
By using the $m$-th moment
\begin{equation}
\Omega_{\mib{q}}^{(m)} =
\int_0^\infty \omega^m A(\mib{q},\omega) d\omega,
\label{eq:moment}
\end{equation}
the excitation energy $\omega_{\rm sw}$ and linewidth $\gamma$ 
of the spin wave excitation are obtained by
\begin{eqnarray}
\omega_{\rm sw} (\mib{q}) &=& 
\Omega_{\mib{q}}^{(1)},
\label{eq:epsilon_q}
\\
\gamma^2 (\mib{q}) &=& 
\Omega_{\mib{q}}^{(2)} - 
(\Omega_{\mib{q}}^{(1)})^2,
\label{eq:gamma_q}
\end{eqnarray}
respectively.
The moments in eq.~(\ref{eq:moment}) can be calculated 
by the magnon self-energy $\Pi_{ij}$.
By using eqs.~(\ref{eq:A_qw}) and (\ref{eq:eigeneq_Pi}),
we obtain
\begin{equation}
\Omega_{\mib{q}}^{(m)} = \frac1N
\sum_{ij} \sum_{k_1 k_2 ...k_{m-1}}
\Pi_{ik_1} \Pi_{k_1 k_2} \cdot \cdot \cdot
\Pi_{k_{m-1} j} 
{\rm e}^{{\rm i}\mib{q} (\mib{r}_i - \mib{r}_j)}.
\label{eq:Omega^m}
\end{equation}
Namely, $\Omega_{\mib{q}}^{(m)}$ is a Fourier transform of $\Pi_{ij}^m$.
This spectral function analysis is valid 
only if the excitation spectrum is single-peaked.
Previous study shows
that this is the case
as long as $\mib{q} \sim 0$
even in the presence of randomness
\cite{Motome2002}.
This will be demonstrated also in Fig.~\ref{fig:disp}
later.
Thus, we can discuss the excitation energy and linewidth
at $\mib{q} \sim 0$ by this spectral function analysis.

%%% excitation energy %%%
First, we analyze the excitation energy (\ref{eq:epsilon_q}).
Hereafter, we assume the electron hopping
only for nearest-neighbor sites in model (\ref{eq:H}) for simplicity.
Then, by eq.~(\ref{eq:Pi}),
$\Pi_{ij}$ has nonzero matrix elements
only for $i=j$ and $j + \mib{\eta}$.
Here $\mib{\eta}$ is a displacement vector to
the nearest neighbor site.
From eqs.~(\ref{eq:epsilon_q}) and (\ref{eq:Omega^m}),
the excitation energy is given by
\begin{eqnarray}
\omega_{\rm sw} (\mib{q}) &=& \frac1N
\sum_i \Big( \Pi_{ii} + 
\sum_{\mib{\eta}} \Pi_{i,i+\mib{\eta}} \
{\rm e}^{{\rm i}\mib{q}\mib{\eta}} \Big)
\nonumber \\
&=&
\sum_{\mib{\eta}} \bar \Pi(\mib{\eta})
({\rm e}^{{\rm i}\mib{q}\mib{\eta}} - 1),
\label{eq:epsilon_q_2}
\end{eqnarray}
where we define the site-averaged quantity
$
\bar \Pi(\mib{\eta}) = \frac1N \sum_i \Pi_{i,i+\mib{\eta}}
$
and use the sum rule (\ref{eq:sumrule}).
The symmetry for the direction of
$\mib{\eta}$ is expected to be recovered 
after the random average, therefore we assume
$\bar \Pi(\mib{\eta}) \equiv - \Lambda$
to be irrespective of $\mib{\eta}$.
Finally, the spin-wave excitation energy is expressed as
\begin{equation}
\omega_{\rm sw} (\mib{q}) = \Lambda
\sum_{\mib{\eta}} (1- {\rm e}^{{\rm i}\mib{q}\mib{\eta}}).
\label{eq:epsilon_q_final}
\end{equation}
Thus, for hypercubic lattices, the spectrum has
the cosine form for small $\mib{q}$ 
even in the presence of disorder.
The quantity $\Lambda$ gives the spin stiffness
since 
\begin{equation}
\omega_{\rm sw} (\mib{q}) \simeq \Lambda q^2
\label{eq:epsilon_smallq}
\end{equation}
in the limit of $q = |\mib{q}| \rightarrow 0$.
The spin stiffness is proportional
to the kinetic energy of the electronic Hamiltonian (\ref{eq:H})
in our analysis
as easily shown by eq.~(\ref{eq:Pi_ii}).

%%% linelidth %%%
Next, we consider the linewidth (\ref{eq:gamma_q}).
Similarly to eq.~(\ref{eq:epsilon_q_2}),
the second moment is written as
\begin{equation}
\Omega_{\mib{q}}^{(2)} 
= \sum_{\mib{\eta}_1 \mib{\eta}_2}
\bar \Pi_{2} (\mib{\eta}_1,\mib{\eta}_2)
(1 - {\rm e}^{{\rm i}\mib{q}\mib{\eta}_1})
(1 - {\rm e}^{{\rm i}\mib{q}\mib{\eta}_2}),
\end{equation}
where
$
\bar \Pi_{2} (\mib{\eta}_1,\mib{\eta}_2)
= \frac1N \sum_i \Pi_{i+\mib{\eta}_1,i}
\Pi_{i,i+\mib{\eta}_2}.
$
In the absence of disorder, we have
$\Pi_{i+\mib{\eta},i} = \Pi_{i,i+\mib{\eta}} 
= - \Lambda$.
Hence the linewidth $\gamma$ becomes zero 
since
$
\Omega_{\mib{q}}^{(2)}
= (\Omega_{\mib{q}}^{(1)})^2.
$
In the presence of disorder, the linewidth becomes finite.
For simplicity, we consider a special $\mib{q}$ 
such as $(q,0,0,\cdot\cdot\cdot)$
on a hypercubic lattice.
Denoting two different cases of $\mib{\eta}_1 = \mib{\eta}_2$ and
$\mib{\eta}_1 = - \mib{\eta}_2$ as
\begin{equation}
  \bar \Pi_{2}(\mib{\eta}_1,\mib{\eta}_2)
  = \Lambda_2 \pm \delta\Lambda_2 
\quad {\rm for} \ \mib{\eta}_1 = \pm \mib{\eta}_2,
\end{equation}
we obtain
\begin{equation}
\Omega_{\mib{q}}^{(2)}
 = 4 \big( \Lambda_{2} (1-\cos q)^2 + 
 \delta\Lambda_{2} \sin^2 q \big).
\end{equation}
In the limit of small $\mib{q}$,
the linewidth is estimated as
\begin{equation}
\gamma^2 (\mib{q}) \simeq 
 4 \delta\Lambda_{2} q^2 + \big( \Lambda_{2} 
 - \frac43 \delta\Lambda_{2} - \Lambda^2 \big) q^4.
 \label{eq:gamma_smallq}
\end{equation}
Therefore, there is a $q$-linear contribution in the linewidth
$\gamma$ in the presence of disorder.

Consequently,
we obtain $\omega_{\rm sw} \propto q^2$ and $\gamma \propto q$
in the limit of $\mib{q} \rightarrow 0$.
This indicates that the spin wave excitation becomes 
incoherent or localized in the vicinity of $\mib{q} = 0$
since $\omega_{\rm sw} < \gamma$.
This incoherent behavior comes from local fluctuations
of the transfer energy of electrons
since the coefficient of the $\mib{q}$-linear term in $\gamma$
is proportional to
\begin{eqnarray}
\delta \Lambda_2 &\propto&
\bar \Pi_2(\mib{\eta},\mib{\eta}) -
\bar \Pi_2(\mib{\eta},-\mib{\eta})
\propto
\sum_i \big( 
\Pi_{i+\mib{\eta},i} - \Pi_{i-\mib{\eta},i} 
\big)^2  \nonumber \\
&\propto& 
\sum_i \big( 
J_{i+\mib{\eta},i} - J_{i-\mib{\eta},i}
\big)^2.
\label{eq:deltaLambda2}
\end{eqnarray}
As increasing $\mib{q}$,
the incoherent regime is taken over by
the marginally-coherent one 
where $\omega_{\rm sw} \propto \gamma \propto q^2$.
We will discuss this crossover in comparison with numerical results
in the following.

%%% numerical results %%%
\noindent
{\it Results} ---
For further understanding, 
we calculate the excitation energy and the linewidth
numerically, and compare them with the above analytic expressions.
In numerical calculations, we obtain the spectral function
$A(\mib{q},\omega)$ 
following the method in ref.~\citen{Motome2002}, and
calculate the moments by the definition (\ref{eq:moment}).
In the following, we discuss two-dimensional cases as an example.
The qualitative feature of the spectrum is almost independent
of the dimension of the system.
The system size is $40 \times 40$ sites and
the electron density is fixed at
$n = \sum_i \langle c_i^\dagger c_j \rangle/N = 0.7$.
The random average is taken for $50$ different realizations of
random configurations.
We examine effects of the on-site and the bond randomness separately.
In both cases, we consider the binary-alloy-type distribution
of randomness as
$
\varepsilon_i = \pm \delta \varepsilon
$
and 
$
t_{ij} = t \pm \delta t,
$
where the sign takes plus or minus in equal probability.
We note that other type distributions, such as Gaussian-type,
do not change conclusions.
As an energy unit, we use the half bandwidth $W=1$
at the ground state for $J_{\rm H} = \delta \varepsilon = \delta t = 0$.
The dimensionless constant $2S$ is set to be unity.

Figure~\ref{fig:disp} shows a typical excitation spectrum
in the presence of randomness.
The gray-scale contrast shows the intensity of the spectrum, namely,
the magnitude of the spectral function $A(\mib{q},\omega)$.
The global structure keeps a portion of the cosine-like one
in the case of pure system
\cite{Furukawa1996},
however, some anomalies appear 
near the zone boundaries.
This is due to the Friedel oscillation of $J_{ij}$
in eq.~(\ref{eq:Pi_ij})
\cite{Motome2002}.
The gray curve shows the excitation energy $\omega_{\rm sw} (\mib{q})$
calculated from eq.~(\ref{eq:epsilon_q}).
As clearly seen,
$\omega_{\rm sw} (\mib{q})$ well describes
the excitation spectrum in the small-$\mib{q}$ regime
where the anomalies are not substantial.
This strongly supports the applicability of the above spectral analysis
with eqs.~(\ref{eq:moment})-(\ref{eq:gamma_q}) in this regime.

The inset in Fig.~\ref{fig:disp} shows the numerical results for
$\omega_{\rm sw} (\mib{q})$ and $\gamma(\mib{q}$)
from $\mib{q} = (0,0)$ to $(\pi,0)$.
They follow $\omega_{\rm sw} \propto q^2$ and $\gamma \propto q$
in the small-$\mib{q}$ region
as predicted in eqs.~(\ref{eq:epsilon_smallq}) and (\ref{eq:gamma_smallq}).
At $\mib{q} \sim 0$, there is the incoherent regime
where $\gamma > \omega_{\rm sw}$ as discussed above.

\begin{figure}
\epsfxsize=8.5cm
\centerline{\epsfbox{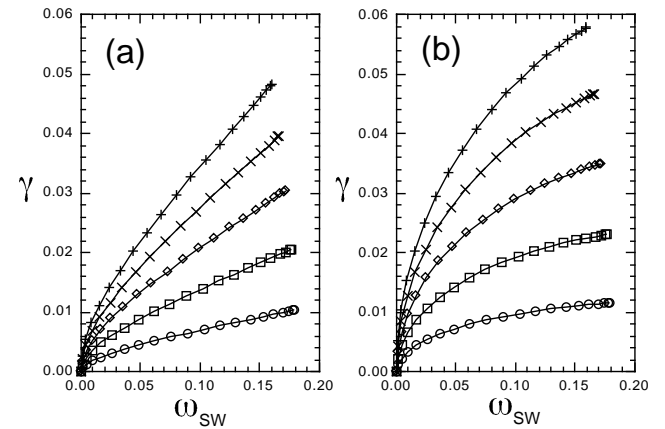}}
\caption{
The linewidth plotted as a function of the excitation energy
in the case of (a) the on-site randomness
$\delta \varepsilon = 0.05, 0.1, 0.15, 0.2, 0.25$
and (b) the bond randomness
$\delta t = 0.025, 0.05, 0.075, 0.1, 0.125$
from bottom to top, respectively.
}
\label{fig:epsilon_vs_gamma}
\end{figure}

In Fig.~\ref{fig:epsilon_vs_gamma},
we compare effects of the on-site and the bond randomness
by plotting the linewidth
as a function of the excitation energy.
For the bond randomness, 
the behavior $\gamma \propto \omega_{\rm sw}^{1/2}$ dominates 
the spectrum as shown in the figure.
On the contrary, for the on-site randomness,
the $\omega_{\rm sw}^{1/2}$ part is observed
only in the small-$\mib{q}$ region, and
the marginally-coherent behavior $\gamma \propto \omega_{\rm sw}$ 
is dominant in the wide region of $\mib{q}$.
Thus, the $\mib{q}$-linear contribution in the linewidth
is more dominant in the case of the bond randomness
compared to the case of the on-site one.
This indicates that the bond randomness tends to
make the spin excitation more incoherent
than the on-site one.

As discussed above
the $\gamma \propto \omega_{\rm sw}^{1/2}$ behavior is 
due to local fluctuations.
Therefore, this different aspect 
between the on-site and the bond randomness
indicates that the bond randomness induces larger
local fluctuations than the on-site one.
This is qualitatively understood as follows.
There are two factors which contribute to 
the local fluctuations
of the transfer energy $(J_{i+\mib{\eta},i} - J_{i-\mib{\eta},i})^2$
in eq.~(\ref{eq:deltaLambda2}).
One is the expectation value $\langle c_i^\dagger c_j \rangle$
and the other is the transfer integral $t_{ij}$.
In the case of the on-site randomness,
we have uniform $t_{ij} = t$, 
hence only $\langle c_i^\dagger c_j \rangle$
contributes to the local fluctuations.
The expectation value $\langle c_i^\dagger c_j \rangle$
is disordered, however,
it may have some correlations due to the Friedel oscillation.
By the Friedel oscillation, the charge density
shows a correlation whose length scale is $\sim 2k_{\rm F}$,
where $k_{\rm F}$ is the Fermi wave number.
On the contrary, in the case of the bond randomness,
$t_{ij}$ fluctuates from site to site independently
in model (\ref{eq:H}).
This may suppress the correlation in $\langle c_i^\dagger c_j \rangle$.
Therefore, we have stronger local fluctuations
and more incoherent spin excitation
in the case of the bond randomness than in the on-site one.

%%% Discussion %%%
\noindent
{\it Comparison with experiments} ---
In real materials, some spatially-correlated or
mesoscopic-scale randomness
may exist due to, for instances,
A-site clustering or twin structure of lattices.
Our results suggest that 
if these kinds of correlated randomness are substantial,
the incoherent regime diminishes and
the marginally-coherent behavior $\gamma \propto \omega_{\rm sw}$
dominates the spectrum
since they may reduce local fluctuations.
Experimentally, the incoherent behavior 
$\gamma \propto \omega_{\rm sw}^{1/2}$ has not been clearly observed yet.
Perring {\it et al.} have reported
an approximately linear scaling of $\gamma \propto \omega_{\rm sw}$ 
in the wide region of the spectrum
in layered perovskite manganites.
\cite{Perring2001}
They observed a significantly large damping, that is,
$\gamma/\omega_{\rm sw} \simeq 0.3-0.5$.
Our results in the marginally-coherent regime 
in Fig.~\ref{fig:epsilon_vs_gamma} (a) are consistent with
these experimental observations.
Therefore, we speculate that
the randomness plays an important role on
the spin dynamics in CMR manganites, and
that correlated or mesoscopic-scale randomness
might be substantial compared to atomic-scale one.

Another possible origin for the linewidth broadening
which survives down to the lowest temperature is
the magnon-electron interaction
\cite{Golosov2000,Shannon2002}.
In this scenario, the linewidth is predicted to be
$\gamma \propto q^{d+3}$ where $d$ is the dimension of the system.
For $d=2$ or $3$, this $q$ dependence is stronger than
in our result (\ref{eq:gamma_smallq}),
and is inconsistent with the above experimental result.

%%%%% Summary %%%%%
%\subsubsection*{Summary}
\noindent
{\it Summary} ---
We have studied the spin dynamics in the double-exchange model
with quenched randomness within the spin wave approximation
in the lowest order of $1/S$ expansion.
We have derived analytic expressions
for the excitation energy and the linewidth
by the spectral function analysis.
Incoherent magnon excitation is found
in the vicinity of the zone center
where the linewidth is proportional to the square-root of
the excitation energy.
As increasing the wave number, 
this incoherent behavior is taken over
by the marginally-coherent one in which 
the linewidth is proportional to the excitation energy.
Atomic-scale randomness enhances the incoherence
through the local fluctuations of the kinetic energy of electrons.
Comparison with experimental results is satisfactory and
suggests an importance of correlated or mesoscopic randomness
in real materials.

%%%%% Acknowledgement %%%%%
This work is supported by  ``a Grant-in-Aid from
the Mombukagakushou''.

%%%%% REFERENCE %%%%%

%%%%% FIGURES %%%%%

%\newpage

%\newpage

\end{document}